\documentclass[runningheads]{llncs}
\usepackage[T1]{fontenc}
\usepackage{graphicx}
\usepackage{hyperref}
\usepackage{color}

\usepackage{amssymb}

\usepackage{comment}

\newcommand\blfootnote[1]{%
  \begingroup
  \renewcommand\thefootnote{}\footnote{#1}%
  \addtocounter{footnote}{-1}%
  \endgroup
}

\begin{document}
\title{Resolving quantitative MRI model degeneracy with machine learning via training data distribution design}
\titlerunning{Resolving qMRI model degeneracy. w/ training data dist. design}
%
\author{Michele Guerreri\inst{1,2}\thanks{m.guerreri@ucl.ac.uk} \and
Sean Epstein\inst{1} \and
Hojjat Azadbakht\inst{2} \and
Hui Zhang\inst{1}}
%
\authorrunning{M. Guerreri et al.}
%
\institute{Department of Computer Science \& Centre for Medical Image Computing, University College London, London, United Kingdom \and
AINOSTICS Ltd., Manchester, United Kingdom}
\maketitle              
\begin{abstract}
Quantitative MRI (qMRI) aims to map tissue properties non‐invasively via models that relate these unknown quantities to measured MRI signals. Estimating these unknowns, which has traditionally required model fitting - an often iterative procedure, can now be done with one-shot machine learning (ML) approaches. Such parameter estimation may be complicated by intrinsic qMRI signal model degeneracy: different combinations of tissue properties produce the same signal. Despite their many advantages, it remains unclear whether ML approaches can resolve this issue. Growing empirical evidence appears to suggest ML approaches remain susceptible to model degeneracy. Here we demonstrate under the right circumstances ML can address this issue. Inspired by recent works on the impact of training data distributions on ML-based parameter estimation, we propose to resolve model degeneracy by designing training data distributions. We put forward a classification of model degeneracies and identify one particular kind of degeneracies amenable to the proposed attack. The strategy is demonstrated successfully using the Revised NODDI model with standard multi-shell diffusion MRI data as an exemplar. Our results illustrate the importance of training set design which has the potential to allow accurate estimation of tissue properties with ML.

\end{abstract}
\section{Introduction}

\blfootnote{This work has been submitted and accepted at the Information Processing in Medical Imaging (IPMI) 2023 conference.}Quantitative magnetic resonance imaging (qMRI) generates images in which voxel intensities are quantitatively related to tissue properties, and promises higher reproducibility and interpretability than conventional qualitative MRI. These advantages come at the cost of requiring parameter estimation (fitting signal models to qMRI data), which underpins some of the key challenges associated with qMRI experiments: (1) high computational cost, due to iterative maximization of a typically non-concave likelihood function, and (2) long acquisition time, due to increased data requirements to support model fitting.

Some qMRI applications, such as microstructure imaging, present a third challenge: multiple parameter combinations generate the same qMRI signal~\cite{RN1}. With conventional fitting approaches this degeneracy induces unstable parameter estimations, preventing the identification of the correct value. In some cases degeneracy can be solved by enriching the acquisition protocol. However, this is not always possible due to hardware limitations or acquisition time constraints.

Supervised machine learning (ML) has been shown to address two of these three challenges. It is orders of magnitude faster than its conventional counterpart \cite{RN2,RN3,RN4}, and can cope with significantly shortened acquisition protocols~\cite{RN2}. The third challenge, signal degeneracy, is as-yet unaddressed, and has been identified as a limitation of existing ML approaches~\cite{RN4}.

However, while we would expect shortened acquisition protocols to introduce or exacerbate model degeneracy, the fact that ML methods can handle these protocols~\cite{RN2} suggests they may be robust to degeneracy in some circumstances. Indeed, Bishop and Roach~\cite{RN5} identified signal degeneracy as a limitation to their seminal ML parameter estimation method. They proposed a simple fix: identifying the degenerate regions of parameter space associated with the same signal, and removing all but one of these regions in the training data. They explicitly resolved degeneracy by imposing a prior on the training data: any one-to-many mappings (one signal to many parameters) are manually reduced to one-to-one. This process is effective but limited by the arbitrary choice of prior: the truncation of the training data is not data-driven.

In this paper we propose a method to resolve signal degeneracy encountered in ML parameter estimation. We present a variation of Bishop and Roach’s approach, where we truncate the training data distribution in a data-driven manner. Our method requires three conditions commonly encountered in qMRI: (i) signal degeneracy is present when using a commonly-used acquisition protocol; (ii) this degeneracy can be resolved by using a super-sampled acquisition protocol not commonly available or accessible; (iii) the super-sampled protocol reveals that all clinically-observed voxels belong to just one of the conventionally-indistinguishable degenerate parameter-space combinations. If these conditions are met, an optimal ML training dataset can be constructed for the commonly-used acquisition protocol by rejecting degenerate parameter combinations which are not observed when using the super-sampled protocol.

This work demonstrates our method with microstructure imaging, but is applicable to any qMRI application for which the above conditions hold. We show that our method resolves degeneracy associated with applying the Revised NODDI model~\cite{guerreri2018revised,guerreri2020tortuosity} to conventional multi-shell diffusion MRI (dMRI) data. This degeneracy can be resolved by acquiring richer tensor-valued diffusion encoding data~\cite{RN9,RN10}. We demonstrate that an ML parameter estimation method, trained within our proposed degeneracy-resolving framework, resolves signal degeneracy when applied to an inherently-degenerate conventional multi-shell acquisition protocol. We use a simple feed forward deep learning (DL) model to represent ML. We use synthetic data for training and test on unseen \textit{in vivo} data.

\section{Theory}

This section first describes the three categories degeneracy can be divided into, then their different impact on ML, and finally the proposed method.

\subsection{Impact of model degeneracy on ML-based parameter estimation}

Traditionally, parameter estimation involves the fitting of a forward model, parameterised by some tissue properties of interest, to measurements. Optimal values for the sought properties are obtained by maximising some likelihood function. With this approach, it is possible that multiple combinations of the properties maximise the likelihood function equally well, a problem known as model degeneracy. While the impact of model degeneracy on conventional parameter estimation is well understood, its effect on ML-based parameter estimation has yet to be fully investigated. Here we characterize this impact by deriving the ML output when a degenerate subset of samples is included during training.

ML-based parameter estimation works by learning a function $f_{\theta}$, parameterised by its learnable parameter vector $\theta$, that can accurately map some measurements  $x \in \mathbb{R}^M$ to the corresponding tissue properties $y \in \mathbb{R}^K$.  Given a set of $N$ exemplar pairs $[x_i,y_i]$, $i=1,...,N$, as the training samples, ML learns the optimal $\hat{\theta}$ that can best approximate the required mapping.  Training algorithms generally minimize the mean squared error (MSE) between the predictions and the corresponding labels for the training set:
\begin{equation}
E_{\theta}= \frac{1}{N} \sum_{i=1}^{N}\sum_{k=1}^{K}[f_{\theta}^k(x_i)-y_i^k]^2 {\mbox .}
\end{equation}

Importantly, for any given input vector $x$, $f_{\theta}$ generates only one output vector $y$; this function can only represent one-to-one or many-to-one mappings.  Degenerate, multi-valued correspondences cannot be represented~\cite{RN5}.

Without loss of generality, we will assume the degenerate subset of samples is the first $n$ samples of the training set.  For this subset, each sample has a distinct output vector but has the same input which we will denote as $x_d$:
\begin{equation}
\forall i,j \in \{1,2,...,n\}, x_i = x_j = x_d\, \mbox{but} \, y_i \neq y_j {\mbox .}
\end{equation}
The MSE loss can then be decomposed into one term involving the degenerate subset and one with the rest of the samples, such that
\begin{equation}
E_{\theta} = \frac{1}{N} \sum_{i=1}^{n} \sum_{k=1}^{K} \left[ f_{\theta}^k (x_d) - y_i^k \right]^2 + \frac{1}{N} \sum_{i=n+1}^{N}\sum_{k=1}^{K} \left[ f_{\theta}^{k} (x_i) - y_i^k\right]^2 {\mbox .}
\end{equation}

If we assume the parameterization of $f_\theta$ is sufficiently flexible such that each of the two MSE terms can be minimized independently, the trained function $f_{\theta}$ will minimize the term involving the degenerate subset $E_{\theta}^d$ when
\begin{equation}
0 = \frac{\partial E_\theta^d}{\partial \theta} = \frac{2}{N} \sum_{k=1}^K \frac{\partial f_\theta^k(x_d)}{\partial \theta} \left[ n f_\theta^k(x_d) - \sum_{i=1}^n y_i^k\right] {\mbox ,}
\end{equation}
which implies that
\begin{equation}
\forall k \in \{1, 2, ..., K\}, f_{\theta}^k (x_d) = \frac{1}{n}\sum_{i=1}^{n}y_i^k {\mbox .}
\end{equation}
This shows in the presence of degenerate samples during training, the learned function will map the degenerate signal $x_d$ to the empirical mean of the tissue properties over the degenerate subset. This highlights the significant role the training data distribution can play in determining the learned mapping: the empirical mean may be tuned, by choosing a suitable training data distribution, to produce the desired mapping.

Crucially, the impact of this result depends on the nature of the degeneracy which can be broadly classified into three categories: (1) degeneracy that genuinely obscures biological differences, i.e. different underlying tissue properties give rise to the same signal; (2) degeneracy that potentially obscures biological differences, but only one non-degenerate case is observed in real data; (3) degeneracy that corresponds to no biological differences.

With Type (1) degeneracy impossible to resolve and Type (3) irrelevant, Type (2) degeneracy is of particular interest as it can be resolved with ML. Namely, by excluding from the training set the combinations of tissue properties which are degenerate, but which are not observed in real data, at inference time the ML framework should be able to correctly estimate the combination which is typically observed.

\subsection{Proposed method to solve degeneracy}
Here we describe the proposed procedure to resolve Type (2) degeneracy based on the observations above. First, we describe the two conditions necessary for the method to be feasible, given a qMRI model and a standard acquisition protocol: (i) the presence of a degeneracy when fitting the model to the data acquired with the standard protocol; (ii) the availability of a rich protocol that can resolve the degeneracy and includes the standard protocol as a subset. When both conditions are met, the observed degeneracy can be identified as Type (2).

\begin{figure}
\includegraphics[width=\textwidth]{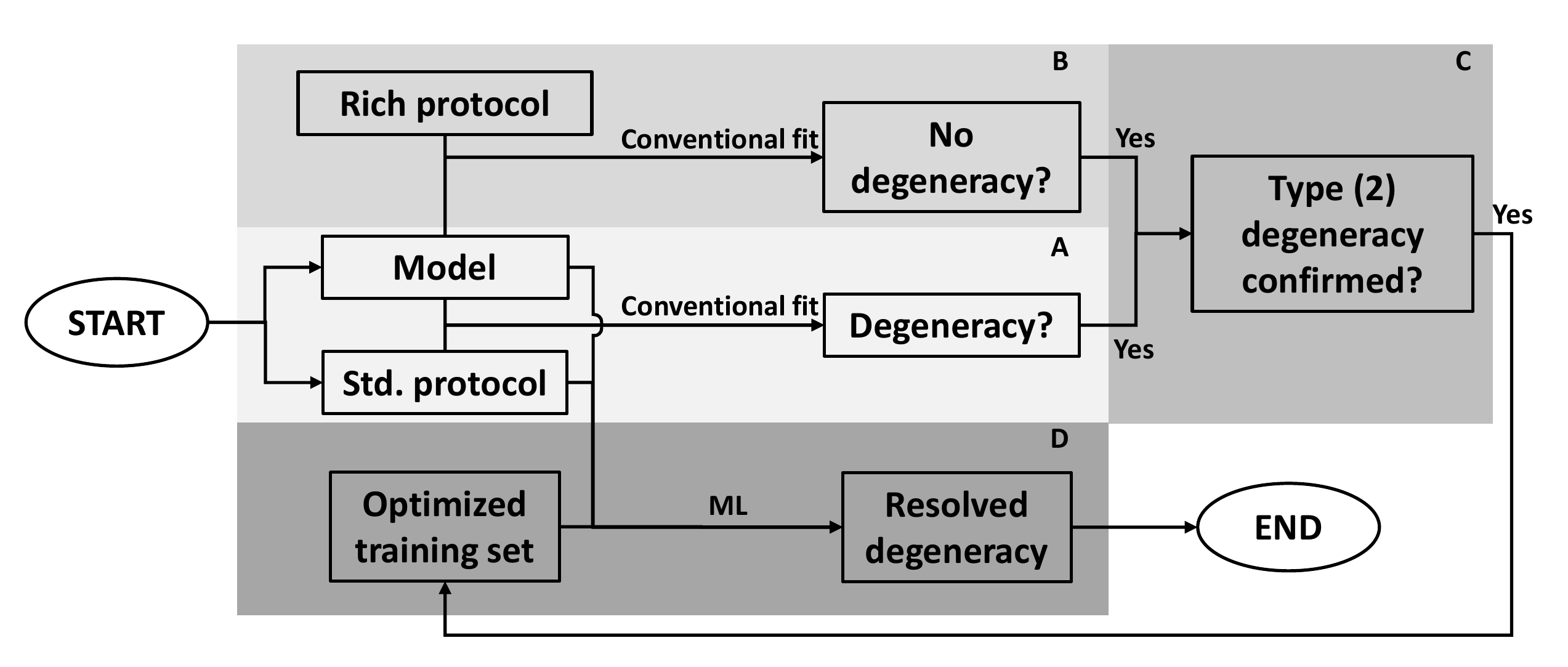}
\caption{Schematic of the proposed method to resolve Type (2) degeneracy: A. given a model and a standard protocol, the presence of degeneracy is assessed via conventional fit; B. a rich protocol exists to assess whether degeneracy can be resolved; C. if both A. and B. are verified, degeneracy is confirmed as Type (2); D. if so, an optimized training set can be constructed removing unobserved degeneracies and used for ML training.} \label{fig1}
\end{figure}

The proposed method comprises four steps as summarized in Figure~\ref{fig1}: First, for a given qMRI model and the standard protocol, establish the presence of degeneracy using the conventional fitting approach on real data acquired with the standard protocol (Panel A). Second, confirm the absence of degeneracy for the data acquired with the rich protocol using the same conventional fitting approach (Panel B). Third, establish the degeneracy is of Type (2) (Panel C). Fourth, construct an optimized training set by excluding all the degenerate parameter combinations which are not observed in Step two. The optimized parameter distribution can then be used to generate the corresponding MRI signals for the chosen qMRI model and the standard protocol to train a ML model.

\section{Methods}
This section describes the datasets, implementation and experiments used to demonstrate the proposed method in a real-world application. The qMRI and ML frameworks used to exemplify our approach are detailed. These are followed by a description of the acquisition protocol of the \textit{in vivo} data and the synthesis of simulated data. Finally, we describe the experiments conducted.

\subsection{qMRI and ML frameworks}
\subsubsection{Revised NODDI}
We demonstrate the proposed method with an exemplar microstructure imaging framework: Revised-NODDI~\cite{guerreri2018revised,guerreri2020tortuosity}, a recently proposed variation on NODDI~\cite{RN6}. NODDI is a popular neuroimaging technique~\cite{RN14} that aims to quantify neurite morphology using standard dMRI protocols. Revised-NODDI overcomes some of the limitations of the original version~\cite{RN17}, and is compatible with the more advanced tensor-encoded dMRI protocols~\cite{guerreri2018revised}.

Similar to NODDI, Revised-NODDI models voxel signals as originating from two compartments: free water and tissue. The tissue signal is further divided into two components: intra-neurite and extra-neurite. For each voxel, Revised-NODDI generates four indices related to microstructural tissue properties. These indices are the neurite density index (NDI), which measures the proportion of the tissue component occupied by neurites; the orientation dispersion index (ODI), which measures the directional variability of neurites; the intra-neurite diffusivity ($d_{I}$), which measures the isotropic diffusivity of the neurites; and the free water fraction (FWF), which measures the extent to which the voxel is contaminated by free water.

Revised-NODDI is chosen as the exemplar because, with standard dMRI protocols, the model is known to be degenerate - its four parameters can not be estimated at the same time. This degeneracy is typically circumvented by assuming $d_I$ can be fixed to some {\it a priori} value, which may be suboptimal under pathological conditions. The alternative is to include additional tensor-encoded acquisitions but they are unavailable in most large-scale neuroimaging studies, such as HCP and UK Biobank.

\subsubsection{Deep learning}
We consider a deep neural network (DNN), modeled on Golkov~\cite{RN2}, as an exemplar ML parameter estimation method. The network inputs are conventional multi-shell dMRI data, while the outputs are the Revised-NODDI model parameters. We include a total of five fully connected layers (one input layer, three hidden layers, one output layer). The width of the input layer is determined by the acquisition protocol (explained below). The width of the hidden layers is 150. Each hidden layer uses ReLU for activation. A dropout of 0.1 is inserted before the output layer, which has a width of 4 and uses a ReLU for activation. The total number of learnable parameters is $46054+150n$ where $n$ is the number of input dMRI data.

\subsection{Datasets}
\subsubsection{{\itshape in vivo} data}
We use an \textit{in vivo} data set to determine the presence of model degeneracy and to assess the performance of the proposed method at inference time. We use the \href{https://github.com/filip-szczepankiewicz/Szczepankiewicz_DIB_2019}{publicly available data} described in Szczepankiewicz et al.~\cite{RN18}. The dataset comprises scans from one healthy subject acquired with two acquisition methods: a standard multi-shell scheme, known as the linear tensor encoding (LTE) in the terminology of tensor-valued diffusion encoding~\cite{RN9,RN10} and a non-standard scheme, known as the spherical tensor encoding (STE) which offers orthogonal information to LTE~\cite{szczepankiewicz2016link}. Each encoding type contains 4 b-values, up to a maximum of $2.0$ ms/$\mu$m$^2$. The two schemes are combined to form the rich scheme for our experiments. We refer to this dataset as the “\textit{szcz\_DIB}” dataset. To restrict the analysis to white matter (WM), LTE data with b-values up to $1.4$ ms/$\mu$m$^2$ was used to estimate diffusion tensors voxel-wise and to subsequently estimate their corresponding fractional anisotropy (FA) values. The resulting FA map was then fed into the FMRIB’s Automated Segmentation Tool (FAST)~\cite{RN20} to obtain a WM mask.

\subsubsection{Synthetic data}\label{Synthetic data}
We use synthetic data to train the DNN model. We generate $N=10000$ Revised-NODDI WM parameter combinations from a synthetic parameter distribution derived from Guerreri et al~\cite{guerreri2018revised}.

\subsection{Implementation}
\subsubsection{Establishing the degeneracy for the standard protocol}
The Revised-NODDI model, that includes $d_{I}$ as free variable, is known to be degenerate when it is fit to conventional multi-shell dMRI data using conventional fitting~\cite{RN21}. We confirm this degeneracy in the LTE data of WM from the \textit{in vivo} \textit{szcs\_DIB} dataset, using an adapted version of the \href{http://mig.cs.ucl.ac.uk/index.php?n=Tutorial.NODDImatlab}{MATLAB NODDI toolbox implementation} (MATLAB version R2022a). Presence of degeneracy is observed qualitatively by the multiple clusters apparent in the estimated tissue property distribution and spatially-incoherent appearance of the estimated tissue property maps (See Figure~\ref{fig3}). This is verified quantitatively by observing that the initialization of the fitting process from different starting points leads to different tissue property estimation with comparable likelihood for the same voxels, confirming the existence of multiple equivalent local maxima in the likelihood landscape.

\subsubsection{Establishing the lack of degeneracy for the rich protocol}
The Revised-NODDI model is known to be non-degenerate when it is fit to data with both LTE and STE~\cite{guerreri2018revised,guerreri2020tortuosity}. We verify this lack of degeneracy associated with the \textit{in vivo} \textit{szcz\_DIB} dataset by fitting WM voxels acquired using LTE and STE together. We use the same fitting approach as above and qualitatively compare the estimated tissue property distribution and the tissue property maps with the ones obtained using LTE data alone (See Figure.~\ref{fig3}).

\subsubsection{Establishing the degeneracy as Type (2)}
This follows directly from above.

\subsubsection{Training datasets generation}
\begin{figure}
\includegraphics[width=\textwidth]{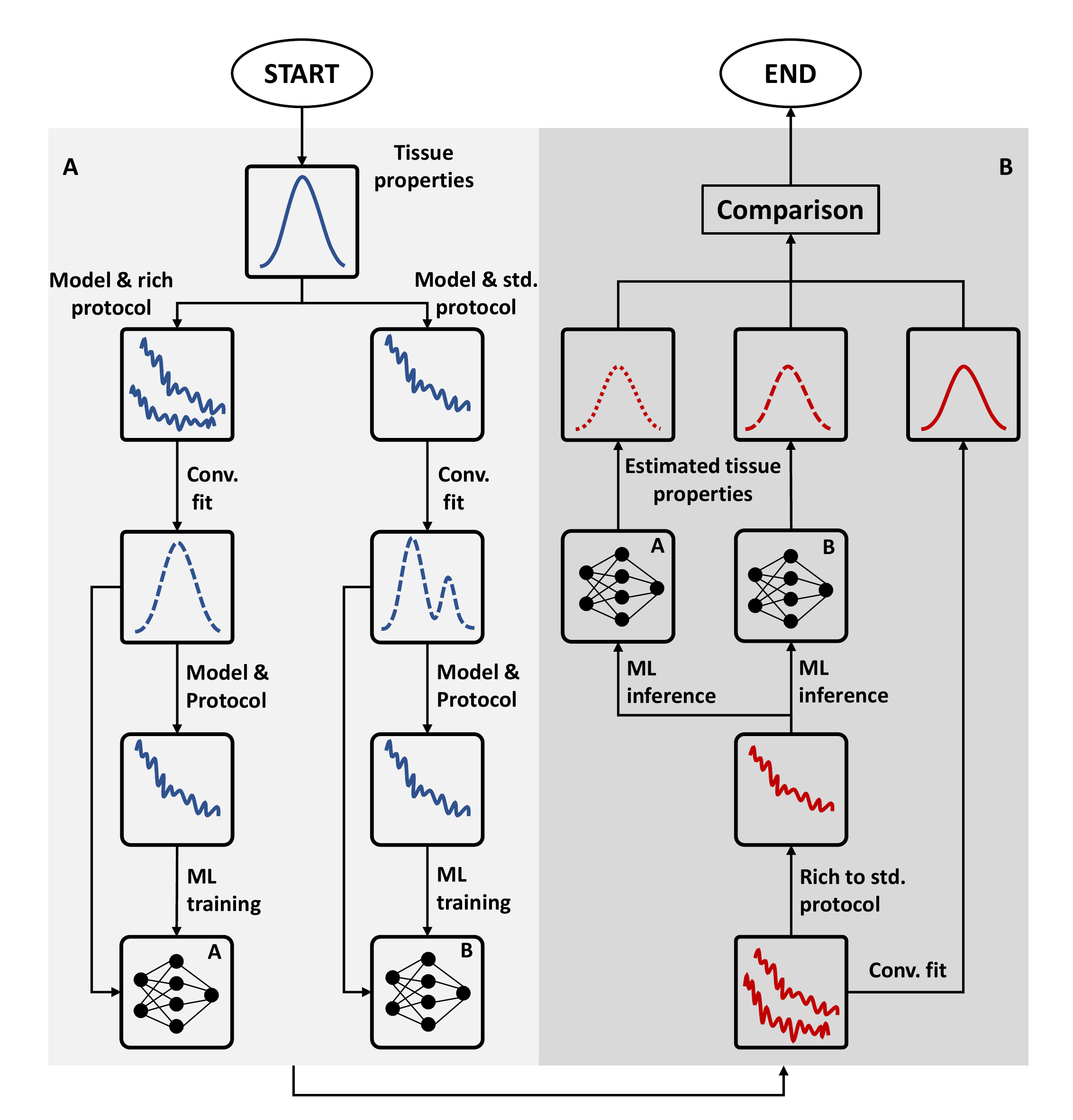}
\caption{Schematic of the experimental design used to demonstrate the proposed method. Panel A: training sets generation and machine learning (ML) training. Two sets are generated which include or not include degenerate tissue properties. Panel B: assessing the effect of different training sets on unseen data.} \label{fig2}
\end{figure}

We generate two training datasets to assess whether a non-degenerate training distribution can be used to resolve signal model degeneracy. This process is summarized in Figure \ref{fig2}.

Both training datasets are generated by initially drawing from the synthetic parameter distribution described above (See Section~\ref{Synthetic data}). The drawn parameters are used to synthesise dMRI signals, using either the \textit{szcz\_DIB} standard protocol (LTE alone) or the rich protocol (LTE+STE), and adding Rician noise that matches the real data (SNR=30). Parameter estimates with the conventional fitting is then obtained for the data simulated with each protocol. By construction, the resulting parameter distribution from the LTE data is degenerate while that from the LTE+STE data is not degenerate. These distributions are subsequently used to generate noisy dMRI signals, using the \textit{szcz\_DIB} LTE protocol, and used to train two DNN models. Data are split 80/20 between training and validation.

\subsubsection{DNN training}
Network training is performed using an Adam optimizer with learning rate = 0.001. A maximum of 1000 epochs is set with an early stopping of 10 epochs.

\subsection{Experiments}
We compare the parameters estimated via the two trained DNN models on the \textit{in vivo} data. We expect to observe the estimations from the network trained via the set with degeneracy to be more biased toward the spurious parameter distribution not observed with the rich protocol.

\section{Results}

\begin{figure}
\includegraphics[width=\textwidth]{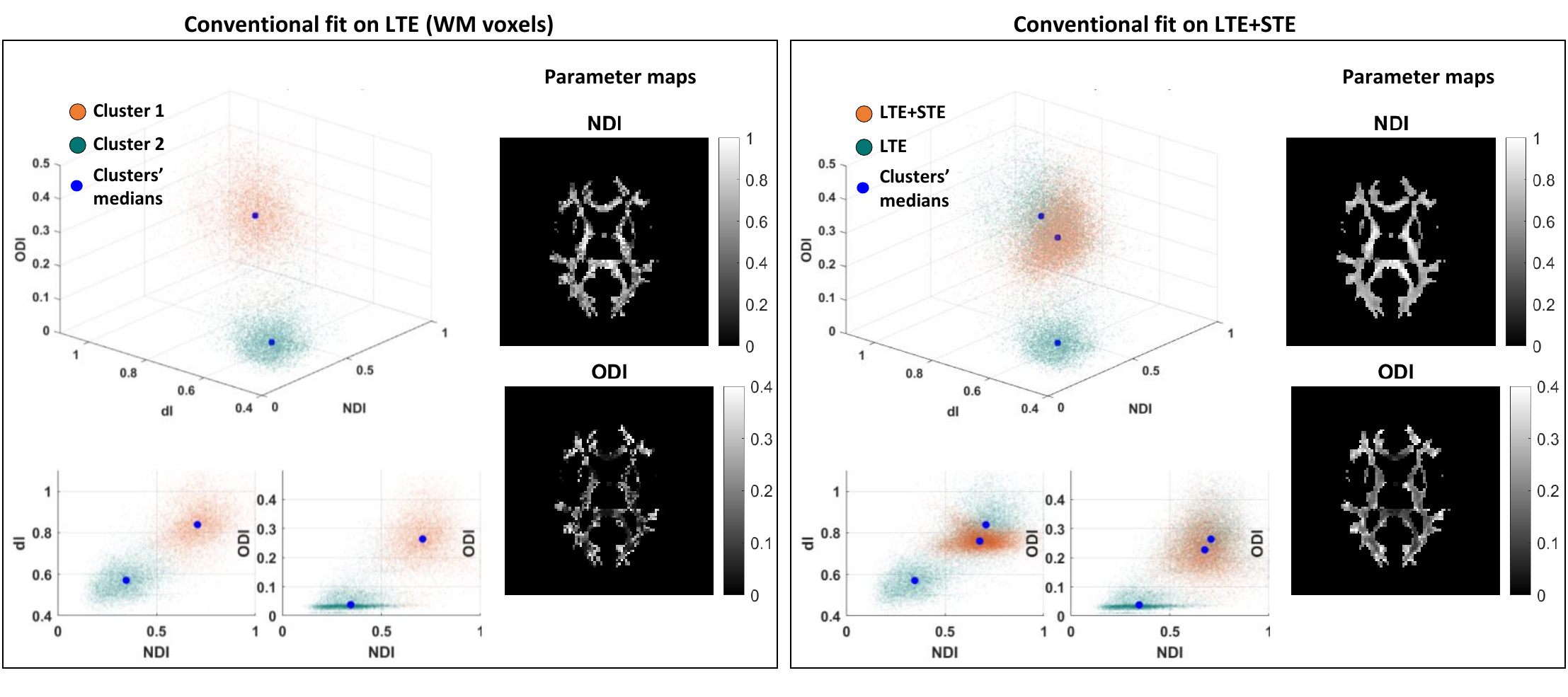}
\caption{
Type 2 degeneracy assessment for Revised-NODDI with conventional linear tensor encoding (LTE) as the standard protocol and including additional spherical tensor encoding (STE) as the rich protocol. Left: tissue properties of Revised-NODDI obtained via conventional fit on LTE \textit{in vivo} data from white matter (WM). The neurite density index (NDI) and orientation dispersion index (ODI) maps appear 'noisy'. In the tissue-property space the voxels appear to organize into two distinct clusters. Together, these observations demonstrate the presence of degeneracy. Right: the degeneracy is resolved using the rich protocol. The non-degenerate tissue properties overlap with only one of the two clusters found with LTE data alone, confirming the observed degeneracy of Type (2).
} \label{fig3}
\end{figure}

Figure~\ref{fig3} shows qualitatively that the degeneracy observed by fitting the Revised-NODDI model to \textit{in vivo} conventional multi-shell LTE data in WM voxels is of Type (2). Left panel shows the estimated tissue properties form two distinct clusters in the tissue-property space. One cluster has higher values of $NDI$, $d_I$ and $ODI$ than the other. Using different starting points to initialize the conventional fit results in different tissue property estimation with comparable likelihood values, which confirm the presence of degeneracy (results not shown). Right panel of the figure shows the degeneracy can be resolved by fitting the model to a richer data set that includes STE. Crucially, the estimated tissue properties from the rich protocol overlap with only one of the two clusters derived from the standard protocol, suggesting that the values in the other cluster is not observed in real data, thus confirming the observed degeneracy is of Type (2).

\begin{figure}
\includegraphics[width=\textwidth]{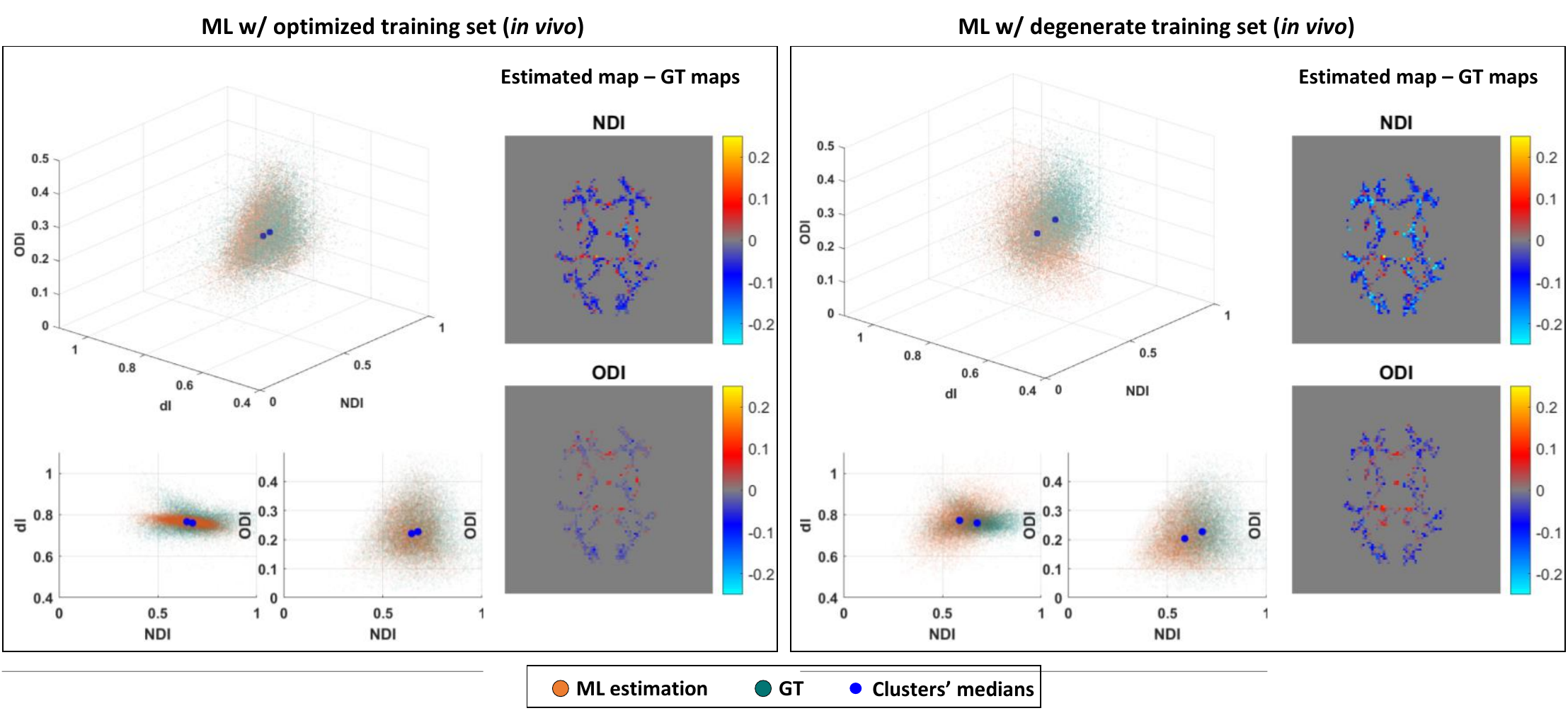}
\caption{
Comparison of estimated tissue property distributions from different machine learning (ML) training strategies. Left: the estimated properties are obtained using a training set which includes only non-degenerate tissue properties. Right: the estimated properties are obtained using a training which includes degenerate tissue properties. The estimated properties are overlaid on top of the ground-truth (GT) values obtained with the rich protocol (Figure~\ref{fig3}). NDI and ODI difference maps between estimated and GT values are also reported.
} \label{fig4}
\end{figure}

Figure~\ref{fig4} shows that training an ML model that excludes the subset of degeneracy parameters not observed in real data can be used to resolve the degeneracy. Left panel shows the tissue properties estimated using a non-degenerate training set, although slightly biased, are close to the parameter distribution derived from the rich protocol with conventional fitting (which we treat as our ground truth). On the other hand, Right panel shows the tissue properties estimated via an ML model that includes the degenerate parameter combinations during training are more biased. The direction of the bias points towards the mean value of the degenerate combinations, as expected from the analysis in the Theory section. The root-mean-squared-errors shown in Table~\ref{tab1} confirm these observations quantitatively.

\begin{table}[ht]
\caption{Root mean squared errors of the estimated tissue properties via machine learning. We compare the estimation error from using training strategies which does or does not include degeneracy. C1(2) for Cluster 1(2) (See Figure~\ref{fig3}).}\label{tab1}
\begin{tabular}{p{1.4cm}p{1.05cm}p{1.05cm}p{1.05cm}p{1.05cm}p{1.05cm}p{1.05cm}p{1.05cm}p{1.05cm}p{1.05cm}}

\multicolumn{1}{p{1.4cm}}{\bfseries Params.} & \multicolumn{3}{p{3.15cm}}{\bfseries Conventional \mbox{}fitting} & \multicolumn{3}{p{3.15cm}}{\bfseries Training set w/ \mbox{}degeneracy} & \multicolumn{3}{p{3.15cm}}{\bfseries Training set w/o degeneracy} \\
\hline
 & C1  & C2 & All & C1  & C2 & All & C1  & C2 & All \\
\cline{2-5}\cline{6-10}
NDI & 0.08 & 0.33 & 0.23 & 0.10 & 0.15 & 0.12 & 0.06 & 0.07 & 0.06 \\
$d_I$ & 0.01 & 0.20 & 0.17 & 0.08 & 0.08 & 0.08 & 0.04 & 0.05 & 0.04 \\
ODI & 0.03 &  0.02 & 0.12 & 0.04 & 0.06 & 0.05 & 0.02 & 0.02 & 0.02 \\
FWF & 0.07 & 0.10 & 0.08 & 0.05 & 0.05 & 0.05 & 0.04 & 0.04 & 0.04 \\
\hline
\end{tabular}
\end{table}

\section{Discussion and Conclusions}
In summary, this paper makes two contributions to the nascent but fast-moving topic of ML-based qMRI parameter estimation. First, we describe a theoretical framework to understand the impact of model degeneracy on this new class of parameter estimation approaches. We show that in general ML-based approaches can not resolve model degeneracy. In contrast to the conventional fitting where model degeneracy leads to unpredictable, thus highly variable, parameter estimates, ML-based approaches instead produce highly consistent but biased estimates. For sufficiently flexible ML models, we show that these biased estimates approach the empirical means of the degenerate training samples, highlighting the key role training data distribution plays in influencing the trained models.

Second, we propose a categorization of model degeneracies that lead to a data-driven scheme of training data distributions. The scheme builds on the idea  first proposed in the seminal work of Bishop and Roach~\cite{RN5} which showed how model degeneracy may be mitigated by tuning the training data distribution, in particular, by removing a subset of the training samples so that the remaining samples no longer contain degeneracy. However, the strategy they used to select the subset for removal was subjective. The data-driven selection scheme proposed here is inspired by the notion of image quality transfer~\cite{alexander2017image} where information contained in high-quality images is used to enhance lower-quality images. If a degeneracy is observed in a standard protocol but not in a richer protocol that includes the standard protocol as a subset, this information can be leveraged to select the appropriate subset among the degenerate parameters for training. This observation leads naturally to the proposed categorization of model degeneracies.

This strategy of embedding the information derived from richer, non-standard protocols can also be implemented in a conventional fitting framework as a Bayesian prior distribution, as elegantly demonstrated in~\cite{mozumder2019population}. One can consider the present work as an ML-based variant that does not require iterative optimization.

\subsubsection{Acknowledgements} Computing resources and support were provided by AINOSTICS Ltd., enabled through funding by Innovate UK. MG is funded by UKRI under grant MR/W004097/1.

%
%
%
\bibliographystyle{splncs04}
\bibliography{ipmi}




\end{document}